\documentclass[12pt]{article}
\newcommand{\ds}{\displaystyle}
\usepackage{hyperref}
\usepackage{float}
\usepackage[utf8]{inputenc}
\usepackage{fullpage}
\usepackage{fullpage, graphicx,psfrag}
\usepackage{hyperref}
 \usepackage{ulem}
\usepackage[ruled,vlined]{algorithm2e}
\usepackage{color}
\usepackage{natbib}

\def\boxit#1{\vbox{\hrule\hbox{\vrule\kern6pt\vbox{\kern6pt#1\kern6pt}\kern6pt\vrule}\hrule}}

\usepackage{color}
\usepackage{url}
\usepackage{setspace}
\usepackage{amsmath,amsfonts,amssymb,amsthm}
         \usepackage{algorithmic}   
\usepackage{hyperref}
\usepackage{mathtools}

\usepackage{xcolor} 
\usepackage{cancel}

\allowdisplaybreaks

\title{Nonparametric Mean and Variance Adaptive Classification Rule for High-Dimensional Data with Heteroscedastic Variances}
\author{  Seungyeon Oh \thanks{Department of Statistics, Sookmyung Women’s University, Seoul, Korea,  \texttt{statsyoh@sookmyung.ac.kr}}~ and Hoyoung Park \thanks{Department of Statistics, Sookmyung Women’s University, Seoul, Korea,  \texttt{hyparks@sookmyung.ac.kr}}}
\date{}

\begin{document}
\maketitle
	\begin{abstract}   


In this study, we introduce an innovative methodology aimed at enhancing Fisher's Linear Discriminant Analysis (LDA) in the context of high-dimensional data classification scenarios, specifically addressing situations where each feature exhibits distinct variances. Our approach leverages Nonparametric Maximum Likelihood Estimation (NPMLE) techniques to estimate both the mean and variance parameters. 
By accommodating varying variances among features, our proposed method leads to notable improvements in classification performance. 
In particular, unlike numerous prior studies that assume the distribution of heterogeneous variances follows a right-skewed inverse gamma distribution, our proposed method demonstrates excellent performance even when the distribution of heterogeneous variances takes on left-skewed, symmetric, or right-skewed forms.
We conducted a series of rigorous experiments to empirically validate the effectiveness of our approach. The results of these experiments demonstrate that our proposed methodology excels in accurately classifying high-dimensional data characterized by heterogeneous variances. 

 \bigskip 
 


\end{abstract}

\section{Introduction}
Recently, researchers have made numerous attempts to obtain new insights by classifying high-dimensional data into two groups. In the context of dealing with this issue, Fisher's Linear Discriminant Analysis (LDA) is a widely used and acknowledged method in various practical applications. However, the high dimensionality of the data occasionally causes immense difficulties in utilizing this method. Specifically, the challenges commonly occur in the following settings. We consider the binary classification using the data, which consists of $p$-dimensional features extracted from $n=n_1 + n_2$ samples. Here, the feature vector and labeling variable of the $i$-th observation are denoted as $\boldsymbol{X}_i = (X_{i1}, \ldots, X_{ip})^T$ and $Y_i \in \{1,2\}$, respectively. 

\begin{equation}
\begin{aligned}
   & \boldsymbol{X_{i}} \mid \left\{Y_{i}=1; \boldsymbol{\mu}_1, \boldsymbol{\Sigma} \right\} \quad \overset{iid}{\sim} \quad N_{p}(\boldsymbol{\mu}_1, \boldsymbol{\Sigma}), \quad i=1, \dots , n_1,\\
   & \boldsymbol{X_{i}} \mid \left\{ Y_{i}=2; \boldsymbol{\mu}_2, \boldsymbol{\Sigma}\right\} \quad \overset{iid}{\sim} \quad N_{p}(\boldsymbol{\mu}_2, \boldsymbol{\Sigma}), \quad i=n_1+1, \dots , n_1 + n_2.\\
\end{aligned} 
\label{eq:intro_setting}
\end{equation}

Assuming condition \eqref{eq:intro_setting} and that the prior probabilities of both groups are $\pi_1 = \pi_2 = 0.5$, the following classification rule known as the Bayes rule can be available.
\begin{align}
       \delta_{OPT}(\boldsymbol{X}^{new}) &= \left( \boldsymbol{X}^{new} - \frac{\boldsymbol{\mu}_1 + \boldsymbol{\mu}_2}{2} \right)^T \boldsymbol{\Sigma}^{-1} \left( \boldsymbol{\mu}_1 - \boldsymbol{\mu}_2 \right) - \log \left( \frac{\pi_2}{\pi_1} \right), \label{eq:bayes_rule1} \\ \widehat{Y}^{new} &= \left \{ \begin{array}{rcl}
        1 & \mbox{for} & \delta_{OPT}(\boldsymbol{X}^{new}) \geq 0, \\
        2 & \mbox{for} & \delta_{OPT}(\boldsymbol{X}^{new}) < 0, \end{array} \right. \label{eq:bayes_rule2}
\end{align}
{where $\widehat{Y}^{new}$ corresponds to the labeling variable for the feature vector $\boldsymbol{X}^{new}$.}

It is important to substitute the unknown parameters included in the optimal classification rule \eqref{eq:bayes_rule1} with robust estimators, since it can potentially enhance the performance of the classifier. Unfortunately, in high-dimensional settings, a large $p$ causes inaccurate estimation of the covariance matrix. In addition, High Dimensional Low Sample Size (HDLSS), which means that $p \gg n$, precludes obtaining the inverse matrix of the estimated covariance matrix due to its singularity. These difficulties hinder efforts to employ LDA in the settings. Certainly, as a strategy to circumvent the significant impediments, we can consider deriving the shrinkage estimator of the covariance matrix in order that the singularity problem can be solved. 
Relevant studies include \cite{ledoit_well_2004,bodnar_strong_2014,ledoit_analytical_2020,ledoitquadratic_2022}, and others. 
{Furthermore, recent studies by \cite{park_high_2022_jmva} and \cite{kim_high_2022} have suggested research findings that involve standardizing data based on the estimation of the precision matrix, and the combination of Nonparametric Maximum Likelihood Estimation (NPMLE) or Nonparametric Empirical Bayes (NPEB) for mean vector estimation contributes to the improvement of Linear Discriminant Analysis (LDA).
}
However, since the computational complexity of the classifiers based on these methods explodes depending on $p$, it is unfeasible to apply these methods in situations when $p$ is excessively large.  

To mitigate the computational burden of the above strategy, adding the Independent Rule (IR) assumption to the optimal classification rule can be a feasible approach. Through the assumption that the components of the feature vector are assumed to be mutually independent, the covariance matrix is simplified to $\boldsymbol{D}={\rm diag}(\boldsymbol{\Sigma})={\rm diag}(\sigma_1^2,\cdots,\sigma_p^2)$. Consequently, it drastically reduces the number of parameters to be estimated, and also allows to invert the estimated covariance matrix without a rank-deficient problem. 
{This approach has been widely researched due to its straightforward interpretation and low computational burden, even in high-dimensional settings.}
For instance, the naive Bayes (NB) is one of the straightforward methods, which estimates $\boldsymbol{\mu}_k$ as the sample mean of the $k$-th group and $\sigma_j^2$ as the pooled sample variances of the $j$-th feature.
Specifically, based on Stein's unbiased risk estimate (SURE), \cite{ji_sure_2020} estimates the mean and inhomogeneous variance parameters using parametric maximum likelihood estimation. SURE exhibits satisfactory performance only in the dense cases, where the mean differences between the two groups are significant in quite a few features. This approach was further extended by adopting 
NPMLE 
in \cite{park_high_2022}. The NPMLE-based method is more adaptive to the mean differences structures as it assumes a nonparametric distribution for mean parameters.  Nevertheless, there remains the potential for enhancement in that the method also incorporates the parametric maximum likelihood estimation for variance parameters. 

Aside from these methods, there are some methods that apply the IR assumption and remove the noise features simultaneously. The nearest shrunken centroid (NSC) from \cite{tibshirani_diagnosis_2002} eliminates the uninformative features using the soft thresholding. The features annealed independence rules (FAIR) from \cite{Fan_High_2008} identifies significant features through the two-sample $t$-test for each of the $p$ features. FAIR is characterized by what uses harder thresholding compared to NSC. Moreover, we can also contemplate the approach estimating $\boldsymbol{\beta}^{Bayes} = \mathbf{\Sigma^{-1}} \left( \boldsymbol{\mu}_1 - \boldsymbol{\mu}_2\right)$. \cite{witten_penalized_2011} proposed penalized linear discriminant analysis (PLDA) which estimates $\widehat{\boldsymbol{\beta}}^{PLDA}$ using $l_1$ penalty and fused lasso penalties. However, these three methods have the disadvantage of performing well only in the sparse cases where the mean differences between two groups are significant in a relatively small number of features.

Considering all the preceding methodologies, we aspire to a robust classifier which performs well regardless of structures of data such as a sparsity of mean differences and distributions of mean and variance parameters. 
Our proposed Mean and Variance Adaptive (\textit{MVA}) linear discriminant rule fulfills all the aspects. Since \textit{MVA} employs the NPMLE to estimate both mean and variance parameters, it is subject to fewer constraints in terms of the structure of data compared to the existing methods. Especially, in SURE, the strong assumption that mean parameters follow a normal distribution can lead to its limited effectiveness in specific cases. The semi-parametric method proposed by \cite{park_high_2022} makes up for this weakness by employing the NPMLE for mean parameters. However, there is still a possibility that its performance could be suboptimal in certain scenarios due to the underlying assumption about the distribution of variance parameters. In contrast to the SURE and NPMLE-base methods of \cite{park_high_2022} which make some specific assumptions related to the distribution of mean or variance parameters, we introduce a more versatile model that doesn't confine itself to specific distributions for estimating mean and variance parameters. The flexibility inherent in our proposed approach is anticipated to enhance accuracy, particularly in the context of high-dimensional binary classification. 
 {The rest of the paper is organized as follows. In the subsequent section, we provide a comprehensive explanation of the simultaneous nonparametric maximum likelihood estimation for mean differences and variances, which underlies our \textit{MVA}. We also introduce a series of algorithms to approximate the Bayes rule. Section \ref{sec:simulation} describes the results of simulation studies conducted on the data 
 with various structures. Section \ref{sec:case_study} compares our \textit{MVA} with the classifiers based on a few prior studies by utilizing gene expression datasets. Lastly, we 
 conclude the paper in Section \ref{sec:conclusion} including concise summaries and 
 outlining 
 future works.}

\section{Methodology}
\subsection{A Simultaneous Estimation Method for Mean Differences and Variances}
In this section, we describe the high-dimensional parameter estimation method to be used in the classification rule proposed in this study. Our proposed classification rule applies the IR assumption so that can alleviate the computational complexity caused by the high-dimensionality. Given the data $\left\{ \left( Y_i, \boldsymbol{X}_i\right) \right\}_{i=1}^n$ satisfying the condition \eqref{eq:intro_setting}, the following $\delta_{I}({\textbf{X}})$ is the optimal classification rule incorporating the IR assumption. 
\begin{eqnarray}
\label{eqn:delta_opt_IR}
\delta_{I}({\textbf{X}}) &=&
\left({\textbf{X}} - \frac{\boldsymbol{\mu}_1 + \boldsymbol{\mu}_2}{2}\right)^{T}
\boldsymbol{D}^{-1}\left(\boldsymbol{\mu}_1-\boldsymbol{\mu}_2\right) 
\coloneqq \sum_{j=1}^p a_j X_{ij} +a_0,
\end{eqnarray}
where {$\boldsymbol{D}={\rm diag}(\sigma_1^2,\cdots,\sigma_p^2)$}. Then, the components to be estimated are represented as follows: 
\begin{eqnarray*}
\label{eqn:a_j}
a_j= \frac{\mu_j}{\sigma_j^2} \quad
	\text{for} \quad \mu_j = \mu_{1j}-\mu_{2j} , \quad a_0= -{2}^{-1}\sum_{j=1}^p a_j(\mu_{1j}+\mu_{2j})
\end{eqnarray*}

Considering the necessity to substitute each $a_j$ with a robust estimator, our primary objective lies in effectively estimating of mean differences $\mu_j=\mu_{1j}-\mu_{2j}$ and variances $\sigma_j^2$ $1\leq j \leq p$ included in \eqref{eqn:delta_opt_IR} from the observed data. Let the observed data $\left\{ \left( Y_i, \boldsymbol{X}_i\right) \right\}_{i=1}^n$, where $\boldsymbol{X}_i = (X_{i 1}, \ldots, X_{i p})^T$, $Y_i=1$ for $i\in\mathcal{C}_1=\{1,\cdots,n_1\}$, and $Y_i=2$ for $i\in\mathcal{C}_2=\{i=n_1+1,\cdots,n_1+n_2=n\}$, and $\mathcal{C}_k$ is a class of indices for the observations belonging to $k$-th group. Assuming the normality of $\boldsymbol{X}_i$, then 
\begin{equation}
	\begin{array}{l}
	X_{i j} \mid\left\{Y_i=k; \mu_{kj}, \sigma_{j}^{2}\right\} \sim N\left(\mu_{kj}, \sigma_{j}^{2}\right), \quad i\in \mathcal{C}_k, \quad k=1,2 
	\end{array} \label{eqn:given_data}
 \end{equation}
 To estimate the unknown components of \eqref{eqn:delta_opt_IR}, we define $X_j$ as a sample mean difference between two groups and $V_j$ as a pooled sample variance as in \eqref{eqn:xj_vj}.
\begin{eqnarray}
    \nonumber X_j&=&X_j^{(1)}-X_j^{(2)}, \text{ where } \displaystyle X_j^{(k)}=       {n_k}^{-1}\sum_{i\in\mathcal{C}_k}X_{ij}, \text{ }k=1,2 \\
    {V_j}&=&\ds \frac{1}{n_1+n_2-2} \left\{\sum_{i\in\mathcal{C}_1}\left(X_{ij}-X_{j}^{(1)}\right)^{2}+\sum_{i\in\mathcal{C}_2}\left(X_{ij}-X_{j}^{(2)}\right)^{2}\right\}, \label{eqn:xj_vj}
\end{eqnarray}
More precisely, we adopt the subsequent Bayesian hierarchical model for the observed $p$ independent pairs of $\left(X_j, V_j\right)$, $1\leq j \leq p,$ as follows: 
\begin{eqnarray}
	\allowdisplaybreaks
	X_j|\mu_j,\sigma_j^2 &\overset{ind}{\sim}& N\left(\mu_j,
	{ \frac{n_1+n_2}{n_1n_2}}\sigma_j^2\right),\quad \mu_j=\mu_{1j}-\mu_{2j}, \label{eqn:dist_X} \\
	{ \frac{(n_1+n_2-2) }{\sigma_j^2}}V_j|\sigma_j^2 &\overset{iid}{\sim}&\chi_{
	{ n_1+n_2-2}}^2,\label{eqn:dist_V}\\
	\mu_j&\overset{iid}{\sim}& G_0\in \mathcal{G}, \label{eqn:main_model_prior_mu}\\
	\sigma_j^2&\overset{iid}{\sim}& F_0 \in \mathcal{F}, \label{eqn:dist_sigma_i^2}
\end{eqnarray}
 
where $\mathcal{G}$ and $\mathcal{F}$ represent classes of all distributions with the support of $(-\infty,\infty)$ and $[0,\infty)$, respectively. Here, we consider $G_0$ and $F_0$ as unknown distributions for location and scale parameters, respectively, and we estimate these distributions by incorporating the observed data structure.  
{Our model assumptions in \eqref{eqn:main_model_prior_mu} and \eqref{eqn:dist_sigma_i^2}, which relax assumptions on the location and scale distributions and aim to estimate them by embracing the data structure, are intended to be more effective in a broader range of situations compared to previous studies that impose assumptions on the distributions of location or scale parameters as in \cite{ji_sure_2020,park_high_2022}}

   

We hereby present a concurrent nonparametric technique for estimating the location parameter, $\mu_j$, and scale parameter, $\sigma_j^2$, $1\leq j\leq p$ as delineated in the model \eqref{eqn:dist_X} $\sim$ \eqref{eqn:dist_sigma_i^2}. Specifically, this paper employs a configuration in which these two parameters are considered independent of each other.


To begin, we derive the optimal Bayes rules for both the location parameter $\mu_j$ and the scale parameter $\sigma_j^2$ for each $j$. Additionally, we introduce a method to approximate these rules. Let $f_{X,V}(X_j,V_j\mid\mu,\sigma^2)$ denotes the joint density function of $X_j$ and $V_j$ given $\mu$ and $\sigma^2$, and $f_{V}(V_j\mid\sigma^2)$ represents the density function of $V_j$ given $\sigma^2$. For $j=1,\ldots,p$, we can characterize these two Bayes rules from model \eqref{eqn:dist_X} $\sim$ \eqref{eqn:dist_sigma_i^2}.

\begin{eqnarray}
        \widetilde{\mu}_j & =&{E}\left(\mu \mid X_j, V_j\right) = \frac{\ds\iint \mu \cdot f_{X,V}(X_j,V_j\mid\mu,\sigma^2)dF_0( \sigma^2) dG_0( \mu)}{\ds\iint f_{X,V}(X_j,V_j\mid\mu,\sigma^2) dF_0( \sigma^2) dG_0( \mu)}
        \label{eqn:Bayes_mj}
\end{eqnarray} 
, and
\begin{eqnarray}
        \widetilde{\sigma}_j^2&=&E\left(\sigma^2\mid V_j\right)=\frac{\ds\int \sigma^2 \cdot f_{V}(V_j\mid\sigma^2) d F_0(\sigma^2)}{\ds\int f_{V}(V_j\mid\sigma^2) d F_0(\sigma^2)}.\label{eqn:Bayes_sigmaj}
\end{eqnarray}   

\subsection{Nonparametric Maximum Likelihood Estimation for $G_0$ and $F_0$ }
\label{sec:npmle_for_G_F}
In most cases, we lack information regarding true distributions $F_0$ and $G_0$, making Bayes' rules \eqref{eqn:Bayes_mj} and \eqref{eqn:Bayes_sigmaj}, inaccessible. Consequently, our primary approach involves approximating Bayes' rules by acquiring estimates of $F_0$ and $G_0$ through the Nonparametric Maximum Likelihood Estimation (NPMLE) \citep{Kiefer_Consistency_1956} and subsequently integrating them into our analysis. 

Consider a parametric family of probability density functions denoted as $\left\{f(\cdot \mid \theta):\theta \in \Theta \right\}$, where $\Theta$ represents a parametric space that is a subset of the real numbers $\mathbb{R}$, and these densities are defined with respect to a dominating measure $d$ on $\mathbb{R}$. Given a distribution $Q$ on $\Theta$, we can define the resulting mixture density as follows:

\begin{eqnarray}
    f_Q(y)\triangleq \int_{\Theta}f(y\mid \theta)dQ(\theta).
    \label{eqn:mixture_density_general_form}
\end{eqnarray}

\citep{Kiefer_Consistency_1956} introduced the Nonparametric Maximum Likelihood Estimator for the distribution $Q(\cdot)$ as the maximizer of the mixture likelihood given a dataset of $p$ observations $y_1,\ldots,y_p$:

\begin{eqnarray}
    \widehat{Q} =  \arg \max _{Q \in \mathcal{M}(\Theta)} \frac{1}{p} \sum_{i=1}^p \log f_Q\left(y_i \right),
    \label{eqn:npmle_general_form}
\end{eqnarray}
where $\mathcal{M}(\Theta)$ represents the set of all probability measure on $\Theta$. For a comprehensive understanding of NPMLE, readers can refer to the well-known study, \cite{Lindsay_Mixture_1995}. 
Therefore, within the NPMLE framework, we can estimate the two distributions, $F_0$ and $G_0$, as follows.

\begin{align}
    \widehat{F}_0 = \operatorname*{argmax}_{F \in \mathcal{F}} \frac{1}{p} \sum_{j=1}^p \left[\log{ \left\{ \int f_{V}\left( V_j \mid \sigma^2 \right) dF(\sigma^2) \right\} } \right],  
    \label{eqn:F_0_hat}
\end{align}
where $ f_{V}\left(V_j \mid \sigma^2 \right)$ is a conditional density function of $V_j$ given $\sigma^2$.
\begin{align}
    \widehat{G}_0 = \operatorname*{argmax}_{G \in \mathcal{G}} \frac{1}{p} \sum_{j=1}^p \left[  \log{ \left\{ \int f_{\widehat{F}_0}\left( X_j, V_j \mid \mu \right) dG(\mu) \right\} } \right],
    \label{eqn:G_0_hat}
\end{align}
where $f_{{F}_0}\left(X_j, V_j \mid \mu \right) = \int f\left(X_j, V_j \mid \mu, \sigma^2 \right)dF_0(\sigma^2)$ and $f\left(X_j, V_j \mid \mu, \sigma^2 \right)$ is expressed as the product of the conditional density function of $X_j$ given $\mu$ and $\sigma^2$ and the conditional density function of $V_j$ given $\sigma^2$.


Although, the convex optimization problem \eqref{eqn:npmle_general_form} is infinite-dimensional, various computationally efficient algorithms have been developed over the years \citep{Jiang_General_2009, Koenker_Convex_2014,Gu_Rebayes_2017}. Following \cite{Lindsay_Mixture_1995,Koenker_Convex_2014,dicker_high_2016, feng_approximate_2016,park_high_2022}, we approximate $\widehat{F}_0$ and $\widehat{G}_0$ with fine grid points support set.
{Concretely, we take a grid of logarithmically equispaced $K$ points $\left\{ \log v_1,\ldots, \log v_K\right\}$ within the interval $\left[\log \underset{1\leq i\leq p}{\min}V_j, \log \underset{1\leq i\leq p}{\max}V_j\right]$. Subsequently, we apply the exponential function to each component of the grid to obtain $\left\{v_1,\ldots,v_K\right\}$.} Also, we consider the $L$ points $\left\{u_1,\ldots,u_L\right\}$ 
{evenly spaced within} the interval $\left[\underset{1\leq i\leq p}{\min}X_i,\underset{1\leq i\leq p}{\max}X_i\right]$. Define $\widehat{\mathcal{F}}_K$ and $\widehat{\mathcal{G}}_L$ as classes of probability distributions supported on the sets $\left\{v_1,\ldots,v_K\right\}$, and $\left\{u_1,\ldots,u_L\right\}$, respectively. We can obtain approximate versions $\widehat{F}_K$ and $\widehat{G}_L$ by substituting $\mathcal{F}$ or $\mathcal{G}$ with $\widehat{\mathcal{F}}_K$ and $\widehat{\mathcal{G}}_L$ in equations \eqref{eqn:F_0_hat} or \eqref{eqn:G_0_hat}. Additionally, these tasks can be easily accomplished using convex optimization algorithms introduced by \cite{Koenker_Convex_2014,Kim_AFast_2020}.

Let
\begin{eqnarray}
    \widehat{F}_K(\sigma^2) &\equiv& \sum_{k=1}^K \widehat{w}_{1,k} I(v_k \leq \sigma^2), \label{eqn:F_K_hat}\\
    \widehat{G}_L(\mu) &\equiv& \sum_{l=1}^L \widehat{w}_{2,l} I(u_l \leq \mu^2)\label{eqn:G_L_hat},
\end{eqnarray}
where $\widehat{w}_{1,k} \geq 0$, $1\leq k \leq K$, $\sum_{k=1}^K \widehat{w}_{1,k}=1$ and $\widehat{w}_{2,l} \geq 0$, $1\leq l \leq L$, $\sum_{l=1}^L \widehat{w}_{2,l}=1$.
Using equations \eqref{eqn:F_K_hat}, \eqref{eqn:G_L_hat}, we obtain viable estimates $ \widehat{\sigma}_i^2$ and $\widehat{\mu}_i$ for $i=1,\ldots,p$ as follows.

\begin{eqnarray}
    \widehat{\sigma}_j^2 & =& \widehat{E} \left( \sigma^2 \mid V_j \right) \nonumber \\
    & =& \frac{\ds\int{\sigma^2f_V(V_j \mid \sigma^2)}d\widehat{F}_K \left( \sigma^2 \right)}{\ds\int{f_V(V_j \mid \sigma^2)}d\widehat{F}_K \left( \sigma^2 \right)} \nonumber \\
    & =& \frac{\ds\sum_{k=1}^K v_k \widehat{w}_{1,k} f_V(V_j \mid v_k)}{\ds\sum_{k=1}^K \widehat{w}_{1,k} f_V(V_j \mid v_k)} \label{eqn:hat_sigma_i}
\end{eqnarray}
\begin{eqnarray}
    \widehat{\mu}_j &=& \widehat{E}(\mu \mid X_j, V_j)\nonumber \\ 
    & = &\frac{\ds\int \int \mu  \widehat{f}_{X,V}(X_j,V_j \mid \mu, \sigma^2) d\widehat{F}_K \left( \sigma^2 \right) d\widehat{G}_L \left( \mu\right)}{\ds\int \int  \widehat{f}_{X,V}(X_j,V_j \mid \mu, \sigma^2) d\widehat{F}_K \left( \sigma^2 \right) d\widehat{G}_L \left( \mu\right)}\nonumber\\
    & =& \frac{\ds\sum_{l=1}^L u_l \widehat{w}_{2,l} \sum_{k=1}^K \widehat{w}_{1,k} f_{X,V}(X_j,V_j \mid u_l, v_k)}{\ds\sum_{l=1}^L \widehat{w}_{2,l} \sum_{k=1}^K \widehat{w}_{1,k} f_{X,V}(X_j,V_j \mid u_l, v_k)}     \label{eqn:hat_mu_i}
\end{eqnarray}

\subsection{Proposed classification rule}

\begin{algorithm}[h]
\caption{Nonparametric Mean and Variance Adaptive
Classification Rule}\label{alg:MVA_alg}
\begin{algorithmic}[1]
 \color{black}
\REQUIRE The $n$ observed data : $\{(Y_i, \boldsymbol{X}_i)\}_{i=1}^n$, $\boldsymbol{X}_i = (X_{i 1}, \ldots, X_{i p})^T$, $Y_i \in \{1,2\}$.
\ENSURE 
\STATE Refer to \eqref{eqn:given_data} and \eqref{eqn:xj_vj} to compute the  sample mean difference between two groups $X_j$ and the pooled sample variance $V_j$ for each $j=1,\ldots,p$. 
\STATE Derive the estimators related to $F_0$ and $G_0$ through the nonparametric maximum likelihood estimation method given by \eqref{eqn:F_0_hat} and \eqref{eqn:G_0_hat}. 
\STATE Substituting $F_0$ and $G_0$ in \eqref{eqn:Bayes_mj} and \eqref{eqn:Bayes_sigmaj} with the estimates of $\widehat{F}_0$ and $\widehat{G}_0$, obtain Bayes estimates $(\widehat{\mu}_j,\widehat{\sigma}_j^2)$ of $(\mu_j, \sigma_j^2)$ for each $j=1,\ldots,p$.
\STATE  Plugging $\widehat{\mu}_j$ and $\widehat{\sigma}_j^2$ into the optimal classification  rule \eqref{eqn:delta_opt_IR}, then \\
$\widehat{a}_j={\widehat{\mu}_j}/{\widehat{\sigma}_j^2}$, $j=1,\ldots,p$, $\widehat{a}_0= -{2}^{-1}\sum_{j=1}^p a_j(X_j^{(1)}+X_j^{(2)})$.
\STATE Classify the new observation $\boldsymbol{X}^{new}$ according to the proposed classification rule below. 
\begin{eqnarray}
        \label{eqn:delta_opt_IR_MVA}
        \delta_{\textit{MVA}}(\boldsymbol{X}^{new}) 
        &=\sum_{j=1}^p \widehat{a}_j X_{ij}^{new} +\widehat{a}_0 - \log \left( \frac{\hat{\pi}_2}{\hat{\pi}_1} \right) \\
        \widehat{Y}^{new} &= \left \{ \begin{array}{rcl}
        1 & \mbox{for} & \delta_{\textit{MVA}}(\boldsymbol{X}^{new}) \geq 0, \\
        2 & \mbox{for} & \delta_{\textit{MVA}}(\boldsymbol{X}^{new}) < 0, \end{array} \right.
\end{eqnarray}
{where $\hat{\pi}_k$ is the proportion of observations belonging to $k$-th group among the all observations.}
\end{algorithmic}
\end{algorithm}

{In this section, we describe our proposed \textit{MVA} classification rule using nonparametric estimation methods for the mean difference and variance parameters introduced in Section \ref{sec:npmle_for_G_F}. The \textit{MVA} rule serves as a classifier that approximates the optimal classification rule  $\delta_{I}({\textbf{X}})$ \ref{eqn:delta_opt_IR} under the IR assumption. The improvement in the estimation of unknown parameters included in $\delta_{I}({\textbf{X}})$ contributes to the enhancement of classification performance. Algorithm \ref{alg:MVA_alg} outlines the construction procedure of the \textit{MVA} rule.}


\section{Simulation Study} \label{sec:simulation}

This section presents a simulation study to assess the performance of \textit{MVA} and compare it with several existing methods. Our main focus is comparing \textit{MVA} with Stein's unbiased risk estimate (SURE) from \cite{ji_sure_2020} and nonparametric maximum likelihood Estimation based method (NPMLE) from \cite{park_high_2022}. 
These methods involve specific assumptions about the distribution of mean or variance values. 
Also, our simulation, features annealed independence rule (FAIR) from \cite{Fan_High_2008}, penalized linear discriminant analysis (PLDA) from \cite{witten_penalized_2011}, naive Bayes (NB), and nearest shrunken centroid (NSC) from \cite{tibshirani_diagnosis_2002} are included. In SURE, we use shrunken estimators towards the grand mean. FAIR is conducted using R-package \texttt{HiDimDA}, PLDA is conducted employing R-package \texttt{penalizedLDA}, and NSC is implemented through R-package \texttt{pamr}. 


{In particular, to compare our \textit{MVA} with SURE from \cite{ji_sure_2020} and NPMLE-based method from \cite{park_high_2022}, which assume a right-skewed distribution for the variance parameters, we set the structures of the variance parameters' distribution as follows. Concretely,} we explore three scenarios related to the skewness of the distribution of the variances in various structures of mean differences. The first scenario involves situations where the distribution of the variances $F_0$ has the long tail on the left side, while the second includes cases where the distribution of the variances $F_0$  is right-skewed. In the last scenario, we evaluate the performance under symmetric distributions for the variances. Section \ref{sec:left} portrays the performance of the models in the first scenario, Section \ref{sec:right} covers the second scenario, and Section \ref{sec:sym} focuses on the last one. This simulation will illustrate our proposed classifier's superior performance in these scenarios. 

\subsection{Data Generation}

{Given $Y_i =k$, all observations are generated from $N_p\left({\boldsymbol{\mu}}_{k}, \boldsymbol{D}\right)$, where ${\boldsymbol{\mu}}_{k}=\left( \mu_{k1},\ldots,\mu_{kp} \right)$ and $\boldsymbol{D}={\rm diag}(\sigma_1^2,\cdots,\sigma_p^2)$. The specific setting for generating the dataset is as follows:}


\begin{enumerate}
    \item The data dimension of $p$ is set to 10,000.

    \item  The sample size of two groups is $n_{1}=n_{2}=125$.
     
    \item 
    {We independently generate $(\boldsymbol{\mu}_k, \boldsymbol{D})$ for each group, and subsequently generate $\boldsymbol{X}_i=(X_{i1}, \ldots, X_{ip}) , \ i=1,\ldots,n_1+n_2$.}
    
    \item For each simulation, we explore both non-sparse and sparse structures of mean differences.
        \begin{itemize}
            \item (Sparse case) \\
                 The first 100 components of $\boldsymbol{\mu}_{1}= \left( \mu_{1j} \right) \ _{1 \leq j \leq p}$ are 1, the remaining components of $\boldsymbol{\mu}_{1}$ are 0 while all of the components of $\boldsymbol{\mu}_{2}= \left( \mu_{2j} \right) \ _{1 \leq j \leq p}$ are 0. To put it differently, $\boldsymbol{\mu}_{1} = (\underbrace{1, \ldots, 1}_{100}, \underbrace{0, \ldots, 0}_{p-100} ) ^T $, $\boldsymbol{\mu}_{2}= \left(0, \ldots, 0 \right) ^T$. 
            \item (Non-sparse case) \\
                The first 100 components of $\boldsymbol{\mu}_{1}= \left(\mu_{1j} \right) \ _{1 \leq j \leq p}$ are 1, the $(p-100)$ remains are generated from $N(0,0.1^2)$. $\boldsymbol{\mu}_{2}$ is set to be the same as the sparse case. 
        \end{itemize}
    
    \item  we split the dataset. For each group, 25 samples are used for training the models and 100 samples are used for evaluation. We repeat 500 times evaluations on independent test datasets in all of the cases.

    
    



       
\end{enumerate}

\subsubsection{Left-skewed case for $F_0$} \label{sec:left}

{In this 
section, we deal with data generated from the various situations where the distribution of the variances $F_0$ is left-skewed. Here, we consider both discrete and continuous distributions for variances. To determine the value of each $\sigma_j^2$ in the discrete distributions for variances, we set $\sigma_{base}^{2}$ as the baseline for $\sigma^{2}_j$, $\delta$ as the proportion of $\sigma_j^2$s with $\sigma_{base}^{2}$ value, and $\Delta$ as the remaining value apart from $\sigma_{base}^{2}$. 
{For instance, consider the scenario where $\sigma_{base}^{2}=1$, $\delta=0.005$, and $\Delta=6$. In this case, for $j=1, \ldots, p$, $\sigma^{2}_{j}$ follows a distribution where $1-\delta=0.995$ proportion of $\sigma^{2}_{j}$ values have $\Delta=6$, and the remaining $\delta=0.005$ proportion of $\sigma^{2}_{j}$ values have a value of 1. This implies that the majority of $\sigma^{2}_{j}$ for $j=1, \ldots, p$ are concentrated on the right side, and as a result, the distribution of variances $F_0$ is considered to be left-skewed.}
In addition, we utilize left-skewed beta distributions for continuous distributions of the variances. To calculate misclassification rates of classifiers in more diverse situations, we consider three different $\sigma_{base}^{2}$ values $(1,2,3)$ and two $\delta$ values $(0.005,0.05)$. The $\Delta$ value is consistently fixed at 6 across all the situations. Moreover, maintaining a scale parameter at 5, we examine three values $(1.5,2.5,3.5)$ for the shape parameter of the beta distributions, denoted as $\beta$. In these situations, we generate $\left\{\sigma_j\right\}_{j=1}^p$ from $\sigma^{2}_j/5 \sim \text{Beta}(5, \beta)$ to accommodate the wider support of the distribution of the variances $F_0$.}

\begin{figure}
    \centering
    \includegraphics[width=1\textwidth]{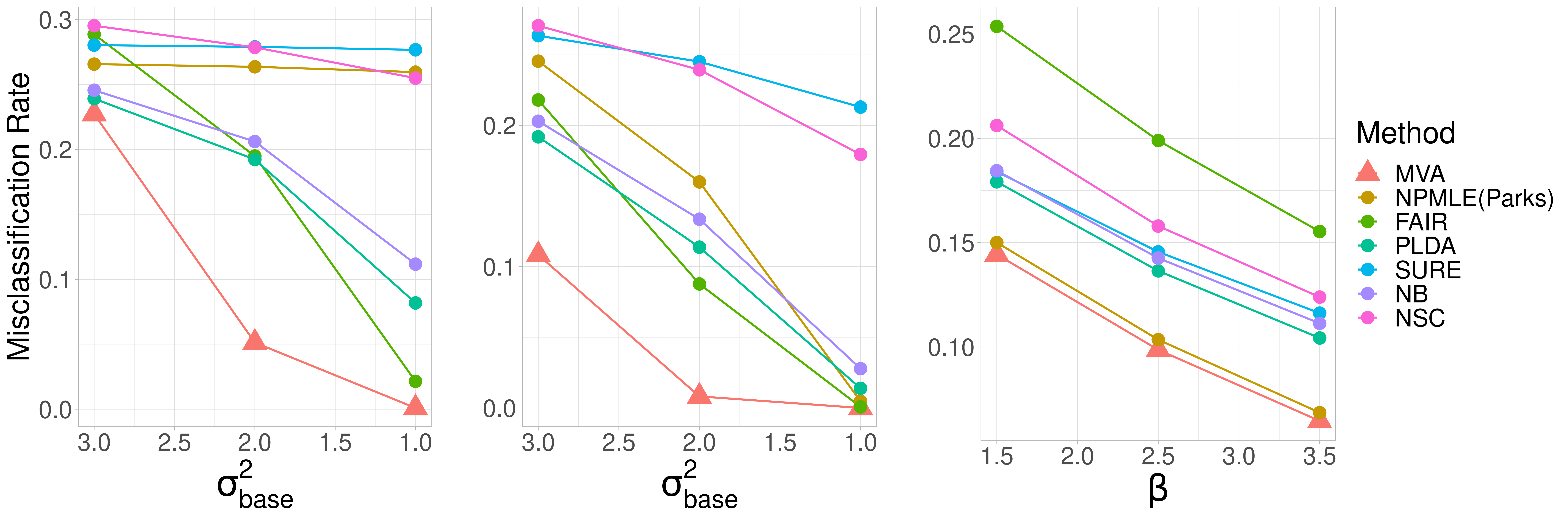}
    \caption{Left-skewed case for $F_0$ with non-sparse setting of mean differences : The left is the situations where $\sigma_{base}^{2}=(1,2,3)$, $\delta = 0.005$, and $\Delta = 6$. The middle is the situations where $\sigma_{base}^{2}=(1,2,3)$, $\delta = 0.05$, and $\Delta = 6$. Lastly, the right is the situations where $\sigma^{2}_j/5 \sim \text{Beta}(5, \beta)$, $\beta=(1.5,2.5,3.5)$.}
    \label{fig:LeftD}
\end{figure}
\begin{figure}
    \centering
    \includegraphics[width=1\textwidth]{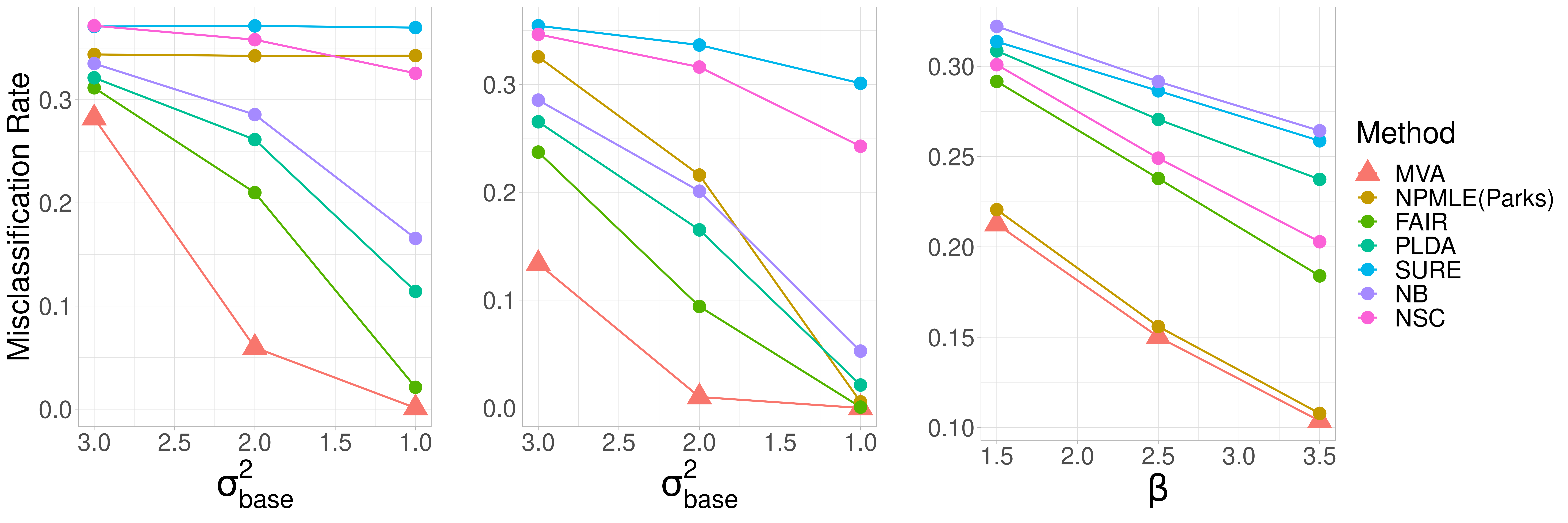}
    \caption{Left-skewed case for $F_0$ with sparse setting of mean differences. In each panel, $F_0$ is set in the same way as in Figure \ref{fig:LeftD}.}
    \label{fig:LeftS}
\end{figure}

Figure \ref{fig:LeftD} and \ref{fig:LeftS} describe the performance of the classifiers in left-skewed case for $F_0$. The left and center of Figure \ref{fig:LeftD} represent the discrete distributions for $\sigma_j^2$, while the right represents the continuous beta distributions for $\sigma_j^2$. Specifically, the left and center panels have different values of $\delta$ set to 0.005 and 0.05, respectively. Hence, the classification problems included in the center panel are considered easier than those in the left panel. The configuration of Figure \ref{fig:LeftS} is the same as Figure \ref{fig:LeftD}, except for the structure of the mean differences. 


In both Figures, \textit{MVA} demonstrates the most superior performance. In the case where the distributions of $\sigma^2_j$ are discrete, the misclassification rates of \textit{MVA} and FAIR are close to zero as the value of $\sigma_{base}^2$ is reduced. On the contrary, some models such as NPMLE(Parks), NSC, and SURE perform poorly in these situations. Especially, NPMLE(Parks) and SURE, which assume a right-skewed inverse gamma distribution for the distribution of $\sigma^{2}_j$, show inferior performance compared to \textit{MVA}. This is because the assumption of the models leads to a conflict with the actual data structure. Our proposed classifier, \textit{MVA}, shows a comparatively smaller misclassification rates due to its flexibility, as it does not make any assumption for the distribution of $\sigma^{2}_j$. In the right panels of Figure \ref{fig:LeftD} and \ref{fig:LeftS}, the skewness of the distribution of $\sigma^2_j$ is mitigated compared to the two previous cases. Nevertheless, no models are able to catch up with the performance of \textit{MVA}. Through these two figures, we show that \textit{MVA} attains the optimal performance among various classifiers in the left-skewed distributions for $\sigma^{2}_j$.

\subsubsection{Right-skewed case for $F_0$} \label{sec:right}

{In this 
section, we examine the performance of the classifiers, including \textit{MVA}, in a few situations where the distribution of variances $F_0$ is right-skewed. As with  
section \ref{sec:left}, we take into account both discrete and continuous distributions for the variances. For discrete distributions of the variances, we explore three different $\sigma_{base}^{2}$ values $(8,9,10)$, two $\delta$ values $(0.005,0.05)$, and a single $\Delta$ value of 6. For continuous distributions of the variances, inverse gamma distributions with a scale parameter of 10 are employed, and three values $(2,4,6)$ for the shape parameter $\alpha$ of the inverse gamma distribution is considered.}

\begin{figure}
    \centering
    \includegraphics[width=1\textwidth]{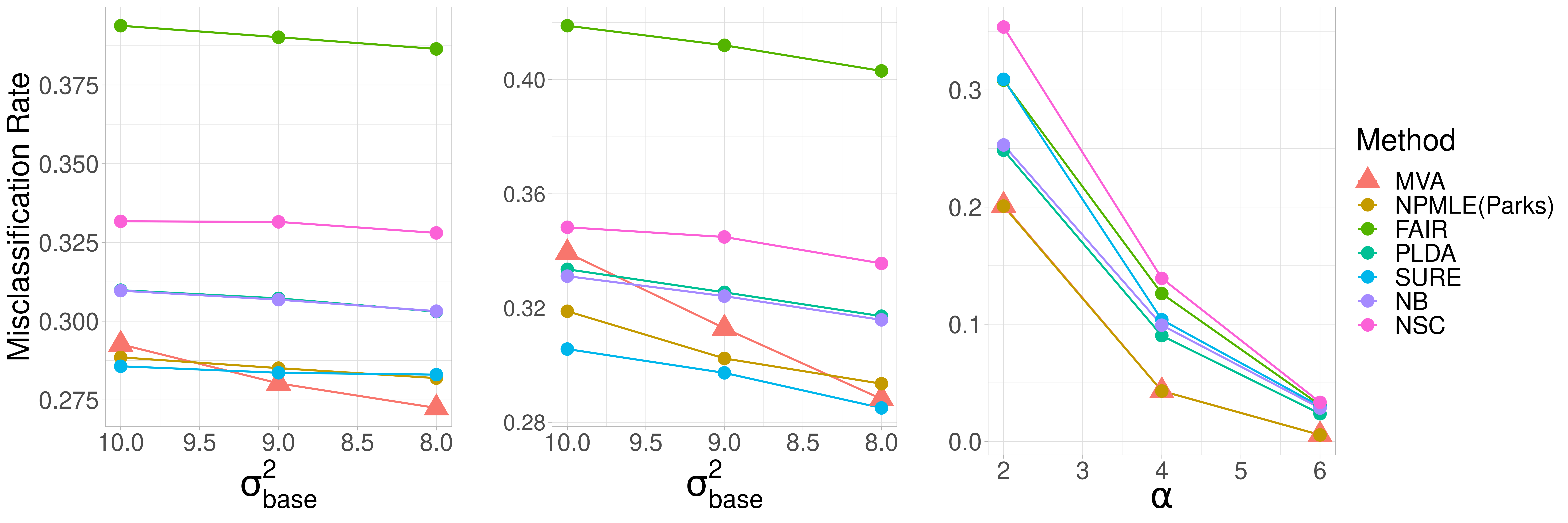}
    \caption{Right-skewed case for $F_0$ with non-sparse setting of mean differences. : The left is the situations where $\sigma_{base}^{2}=(8,9,10)$, $\delta = 0.005$, and $\Delta = 6$. The middle is the situations where $\sigma_{base}^{2}=(8,9,10)$, $\delta = 0.05$, and $\Delta = 6$. Lastly, the right is the situations where $ \sigma^{2}_j \sim \Gamma^{-1}(\alpha, 10)$, $\alpha=(2,4,6)$.}
    \label{fig:RightD}
\end{figure}
\begin{figure}
    \centering
    \includegraphics[width=1\textwidth]{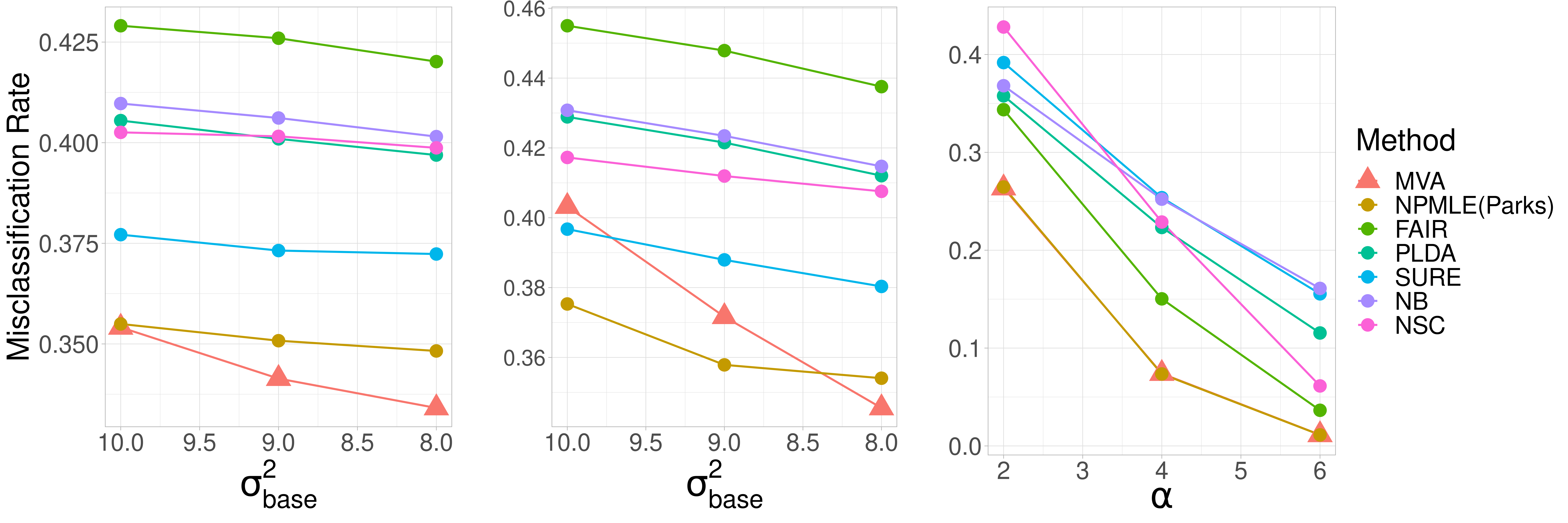}
    \caption{Right-skewed case for $F_0$ with sparse setting of mean differences. In each panel, $F_0$ is set in the same way as in Figure \ref{fig:RightD}.}
    \label{fig:RightS}
\end{figure}

Figure \ref{fig:RightD} and \ref{fig:RightS} present the misclassification rates of the models when the distributions of $\sigma^{2}_j$ are right-skewed. In this case, we anticipate that NPMLE(Parks) and SURE have it over on \textit{MVA} as their assumptions are consistent with the actual forms of the distribution of $\sigma_j^2$. Also, in the left and center panels of Figure \ref{fig:RightD} and \ref{fig:RightS}, $\delta$ is set to 0.005 and 0.05, respectively. These four plots show that NPMLE(Parks), SURE and \textit{MVA} perform well as opposed to the other models. Particularly, NPMLE(Parks) and SURE exhibit outstanding performance at any $\sigma_{base}^{2}$ unlike section \ref{sec:left}. When the value of $\sigma_{base}^{2}$ becomes 10, \textit{MVA} falls behind NPMLE(Parks) and SURE. But it eventually achieves relatively lower misclassification rates than other models when $\sigma_{base}^{2}$ reaches its smallest value. In the right plots of Figure \ref{fig:RightD} and \ref{fig:RightS}, although the distribution of $\sigma^2_j$ is right-skewed, the performance of \textit{MVA} is similar to NPMLE(Parks). As illustrated in those figures, we demonstrate that our \textit{MVA} achieves accomplished performance even when the skewness of the distribution of $\sigma^{2}_j$ is positive.

\subsubsection{Symmetric case} \label{sec:sym}

\begin{figure}
    \centering
    \includegraphics[width=0.7\textwidth]{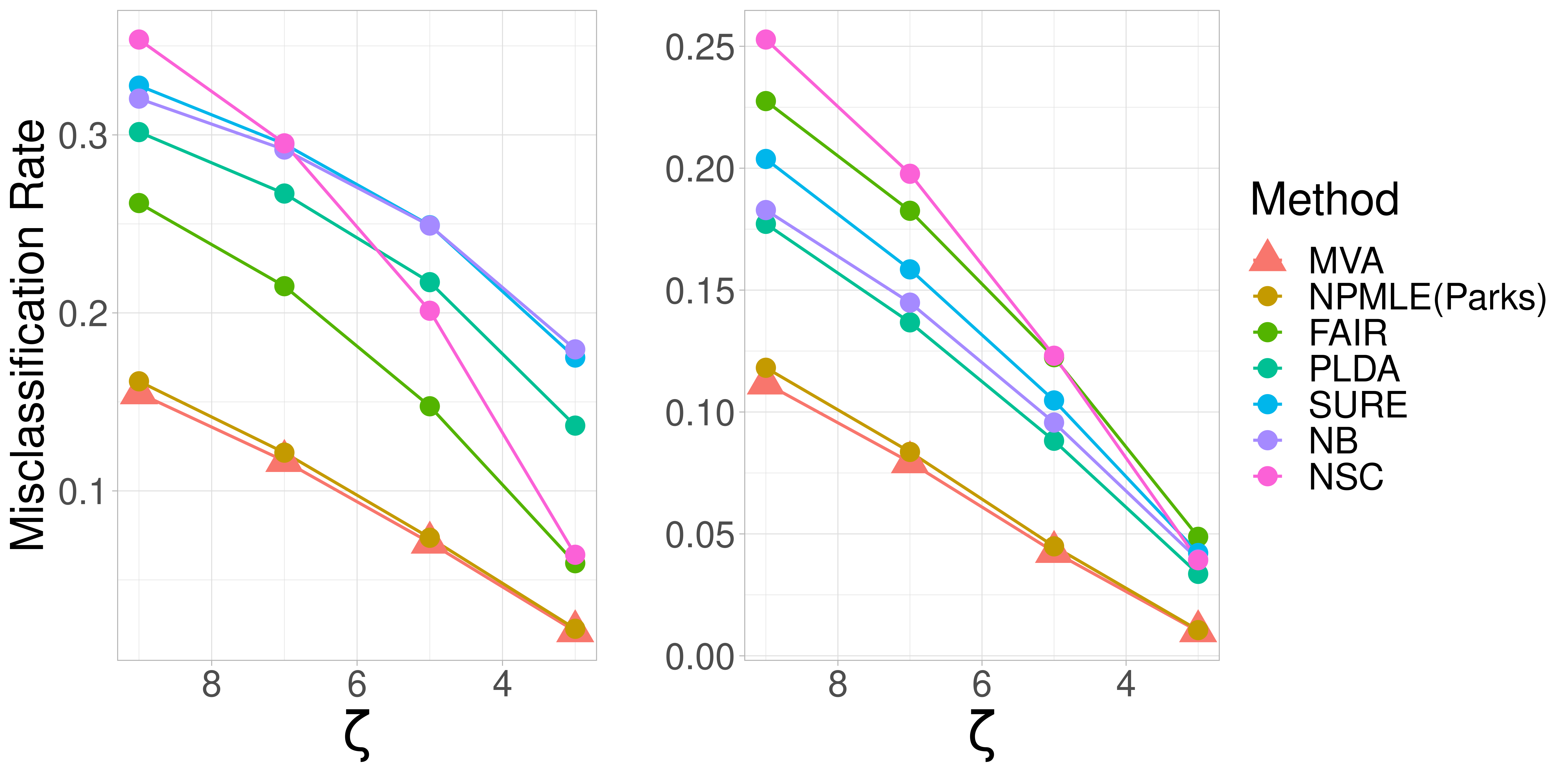}
    \caption{Symmetric case for $F_0$ with both sparse and non-sparse settings of mean differences. : The left is set to sparse structures of mean differences, and the right panel is set to non-sparse structures of mean differences. In both panel, $\sigma^{2}_j \sim U(1, \zeta)$, $\zeta = (3,5,7,9)$.}
    \label{fig:sym}
\end{figure}

{In this 
section, we address the various situations where the distribution of the variances $F_0$ is symmetric, specifically following a uniform distribution. For each situation, we investigate how the performance of the classifiers changes as we decrease the value of $\zeta$, which is the maximum of a uniform distribution 
{with a fixed minimum value set to 1.} In this context, we consider four different values $(3,5,7,9)$ for $\zeta$.}
In Figure \ref{fig:sym}, the results of the simulation are shown when the distributions of $\sigma^{2}_j$ are set to continuous uniform distributions with both non-sparse and sparse structures of mean differences. Here, if $\zeta$ takes relatively larger value, the support of the distribution of $\sigma_j^2$ is wider, and the classification problem in this situation is more challenging. \textit{MVA} and NPMLE(Parks) work well irrespective of the sparseness of the mean differences not only when the support of the distribution of $\sigma^{2}_j$ is relatively broader but also when it is less so. Ultimately, we showed that our proposed \textit{MVA} has promising performance regardless of the skewness of the distribution of $\sigma^{2}$ and the sparseness of the mean differences.

\subsection*{}

\begin{figure}[H]
    \centering
    \includegraphics[width=1\textwidth]{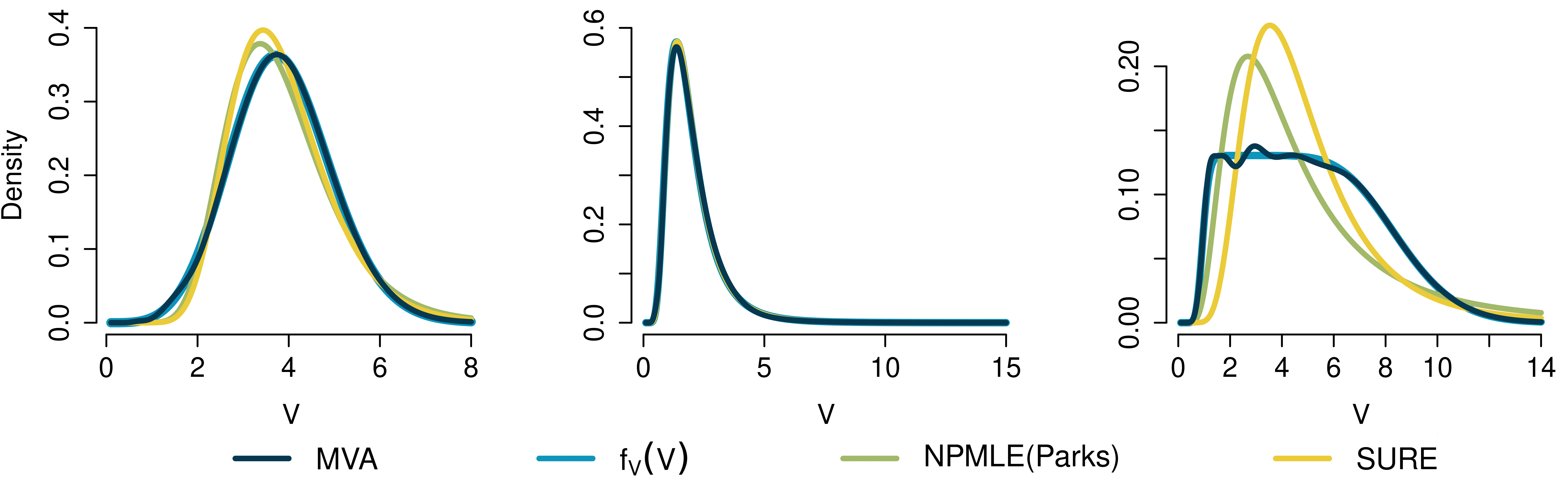}
    \caption{$f_V (V)$ and estimated probability marginal density functions of pooled sample variances in sparse case:  The left is a left-skewed case where $\sigma_j^2/5 \sim \text{Beta}(5,1.5)$, the middle is a right-skewed case where $\sigma_j^2 \sim \Gamma^{-1}(6,10)$, and the right is a symmetric case where $\sigma_j^2 \sim U(1,9)$.}
    \label{fig:spvar_sparse}
\end{figure}

In the above experiments, our proposed model demonstrates stable performance compared to other models in any case. One of the factors that significantly affect the performance of the classifiers is how accurately they estimate mean, and variance values included in the optimal classification rule \eqref{eq:bayes_rule1}. Especially, our model has the advantage of estimating variance values more adaptively than NPMLE(Parks) and SURE. To indirectly confirm that \textit{MVA} estimates the distribution of variance values more flexibly, we need to examine how each model estimates the probability marginal density function of pooled sample variances, utilizing the fact that the pooled sample variance $V_j$ is an unbiased estimator of $\sigma^2_j$. 



Figure \ref{fig:spvar_sparse} compares the estimated probability marginal density functions of pooled sample variances for \textit{MVA}, NPMLE(Parks) and SURE with the corresponding true probability marginal density function, denoted as 
{$f_V(V) = \ds\int f_{V}(V\mid\sigma^2) d F_0(\sigma^2).$} In the left of Figure \ref{fig:spvar_sparse}, where the distribution $F_0$ of $\sigma_j^2$ is left-skewed, \textit{MVA} demonstrates a remarkable adaptability for closely approximating the true distribution of pooled sample variances. However, both NPMLE(Paks) and SURE tend to estimate the $f_V (V)$ slightly deviating from it. In the middle of Figure \ref{fig:spvar_sparse}, which represents one of the right-skewed cases for the distribution $F_0$ of $\sigma_j^2$, all of the models estimate the $f_V (V)$ similarly well. Furthermore, as illustrated in the right of Figure \ref{fig:spvar_sparse}, \textit{MVA} flexibly estimates the $f_V (V)$ by reflecting the actual structure of the data even when the distribution $F_0$ of $\sigma_j^2$ is symmetric. In contrast, NPMLE(Parks) and SURE are constrained by the strong assumption concerning the distribution $F_0$ of $\sigma_j^2$, leading to the failure in accurately estimating $f_V (V)$.

To sum up, our suggested model outperforms the other models except in the situations where the distribution $F_0$ of $\sigma_j^2$ is right-skewed, and few $\sigma_j^2$ have relatively larger values like 10. Particularly, in the left-skewed cases, our classifier exhibits outstanding performance by sharply reducing its misclassification rates as the value of $\sigma_{base}^2$ is decreased. In the right-skewed case, NPMLE(Parks) and SURE present satisfactory performance, and our model is comparable to the two previous models in terms of performance. This tendency remains in both cases where the structure of mean differences is sparse and non-sparse. Consequently, our model fulfills our desire for a robust model that is not influenced by the sparseness of mean differences or skewness of the distribution of variances. As evidence of that, we present a comparison of the estimated probability marginal density of pooled sample variances. Through that, we discover that the robustness of our model is based on adaptive estimation capturing the structure of the actual data. From above the simulation study, we confirm that the classifier based on our devised methodology takes advantage of ensuring effective performance in any case since it estimates mean and variance values without some constraints or assumptions. 

\section{Case Study} \label{sec:case_study}
In this section, we evaluate the performance of \textit{MVA} and several methods by applying them to the four benchmark datasets:
Breast Cancer from \cite{ZHU20073236}, Huntington's Disease from \cite{borovecki_genome_2005}, Leukemia from \cite{golub_molecular_1999}, Central Nervous System(CNS) from \cite{pomeroy_prediction_2002}. 
The Breast Cancer dataset is available at the \url{https://csse.szu.edu.cn/staff/zhuzx/Datasets.html.}, and the rest are provided in R-package \texttt{datamicroarray}. 

Before analyzing the datasets, we carry out the following preprocessing steps. To alleviate the computational burden, we conduct a two-sample $t$-test for each feature and employ a feature screening procedure to identify noisy features, using a significance level of 0.2. These tasks are carried out for all considered datasets. Additionally, we implement min-max scaling on Huntington's Disease and Leukemia to ensure that all the feature's minimum and maximum values are adjusted to 0 and 1, respectively. The details related to the four datasets that are used are presented in Table \ref{tab:Dataset}. 

\begin{table}[t]
    \centering
    \small
    \renewcommand{\arraystretch}{1.5}
    \begin{tabular}{l|c c c c c}
        \hline Dataset & $n$ & $p$ & Classes & Reduced dimension&   \\ \hline Breast Cancer & 97 & 24481 & non-relapse(51), relapse(46) & 6906 \\ Huntington's Disease &  31 & 22283 & control(14), symptomatic(17) & 11455  \\ Leukemia & 72 & 7129 & ALL(47), AML(25) & 3502  \\ CNS & 60 & 7128 & died(21), survived(39) & 1151 \\ 
        \hline
    \end{tabular}
    \caption{Details of the utilized datasets}
    \label{tab:Dataset}
\end{table}

\begin{table}[t]
    \centering
    \small
    \renewcommand{\arraystretch}{1.5} 
    \begin{tabular}{l|c c c c c c c}
        \hline Dataset & MVA & NPMLE(Parks) & FAIR & PLDA & SURE & NB & NSC  \\ \hline
        Breast Cancer & 0.206 & 0.216 & 0.351 & 0.268 & 0.443 & 0.268 & 0.309 \\ 
        Huntington's Disease & 0.000 & 0.032 & 0.065 & 0.065 & 0.065 & 0.032 & 0.065 \\ 
        Leukemia & 0.083 & 0.083 & 0.083 & 0.083 & 0.097 & 0.083 & 0.028 \\
        CNS & 0.200 & 0.217 & 0.383 & 0.233 & 0.217 & 0.217 & 0.250\\
        \hline
    \end{tabular}
    \caption{Misclassification rates of classifiers}
    \label{tab:CaseStudy}
\end{table}

{More specifically, first, 
the breast cancer dataset was employed to predict the relapse status of 97 individuals, resulting in a preprocessed dataset with 6906 genes. Among the samples, only 46 patients experienced a recurrence of breast cancer. 
The second dataset focused on Huntington's disease, aiming to distinguish between 17 symptomatic individuals and 14 in the control group, resulting in a dataset with 11455 genes after preprocessing. 
The third dataset, leukemia data, comprised 47 patients diagnosed with acute lymphoblastic leukemia (ALL) and 25 with acute myeloid leukemia (AML), with the data dimension reduced to 3502 through feature screening. 
Lastly, the CNS dataset is associated with the survival outcomes (died or survived) of 60 patients with central nervous system embryonal tumors, where 21 patients succumbed, and the remaining survived. After feature screening procedures, the dataset retained 1151 genes.
}

\begin{figure}
    \centering
    \includegraphics[width=1\textwidth]{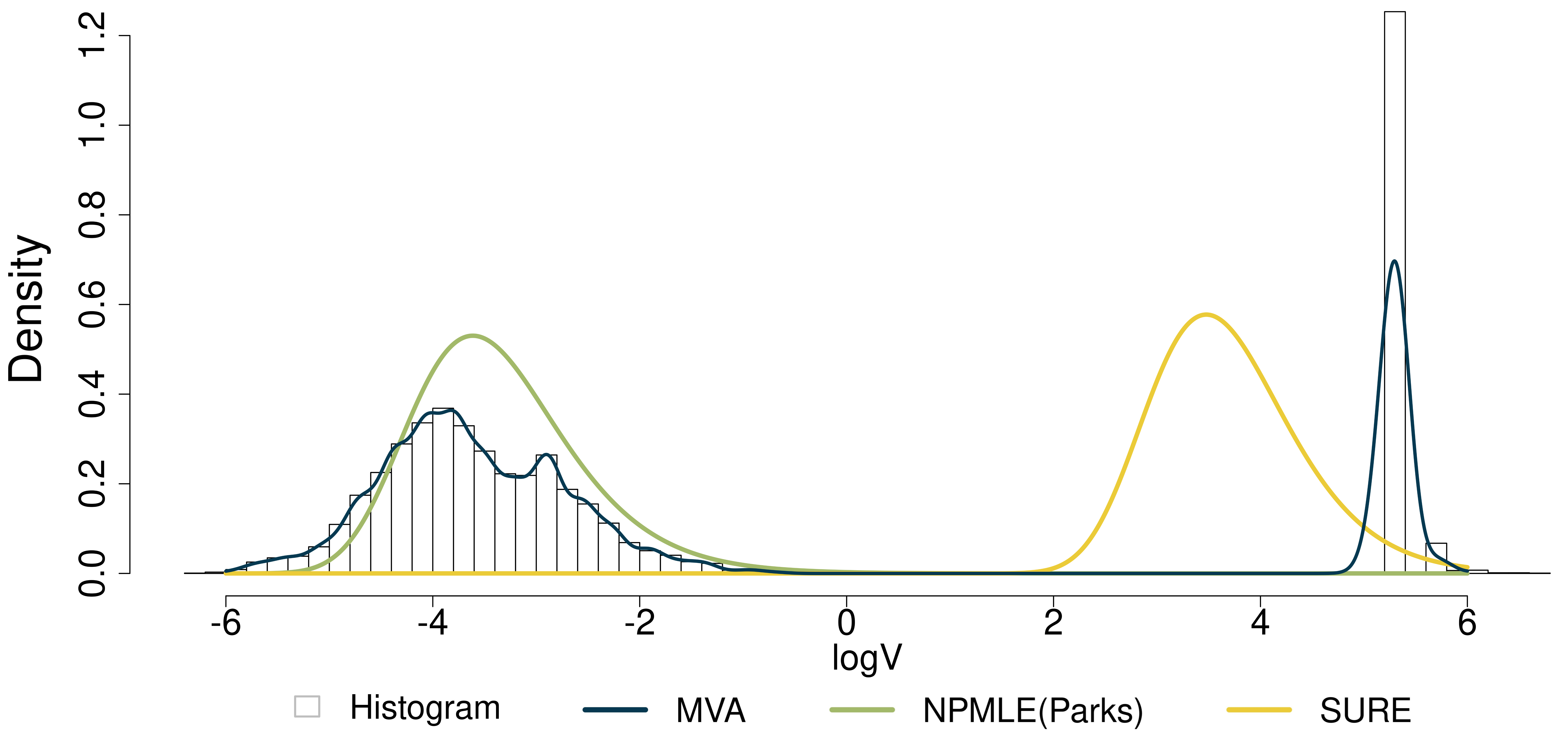}
    \caption{Histogram and estimated probability density functions of log pooled sample variances of the breast cancer dataset}
    \label{fig:spvar_breast}
\end{figure}

The misclassification rates of the models, calculated through Leave-One-Out Cross-Validation(LOOCV), are shown in Table \ref{tab:CaseStudy}. Our proposed model demonstrates satisfactory performance overall for all datasets. As seen in the first row of Table \ref{tab:CaseStudy}, our model exhibits the superior performance for the breast cancer data, while other models, except NPMLE(Parks), show remarkably low performance compared to \textit{MVA}. The results of Huntington's disease data are also found in the second row in Table \ref{tab:CaseStudy}, from which we can see that only our proposed model accurately predicts the status of all the samples. The third row in Table \ref{tab:CaseStudy} presents the misclassification rates of the models for the leukemia data. In this classification problem, our model is not the best in terms of performance, but it still exhibits the performance on par with NSC, which is the optimal model for the data. From the last row of Table \ref{tab:CaseStudy}, our model reveals the top performance in CNS data, and NPMLE(Parks), SURE, and NB have the same performance slightly below that of \textit{MVA}. In contrast, FAIR and NSC have worse performance for this dataset. Consequently, our classifier maintains the robust performance across all the datasets, unlike the classifiers based on the existing methodologies that perform well only on specific datasets.

Additionally, utilizing the analysis of the breast cancer data, we provide evidence supporting the effectiveness of our proposed model in real data analysis. Figure \ref{fig:spvar_breast} particularly aims to show enhanced adaptability of \textit{MVA} in estimating variance values compared to SURE and NPMLE(Parks). We need to remark that the primary difference between our model and the others lies in the assumption related to the distribution of variance values. Since we cannot compare the directly estimated probability density function of the variances for each method, we try to examine which method estimates the distribution of the log pooled sample variances more effectively. In Figure \ref{fig:spvar_breast}, our model effectively detects the bimodality of the distribution represented in the histogram, while the other fails to do so.

In summary, we confirm that our \textit{MVA} has stable performance in most cases since it is based on a more flexible estimation of the parameters included in the optimal classification rule. In other words, if we are given a microarray dataset with noise appropriately removed, we anticipate that our model performs well regardless of the structures of the actual dataset in the context of the high-dimensional binary classification task.

\section{Conclusion} \label{sec:conclusion}

{Our innovative methodology has shown significant promise in enhancing the effectiveness of Fisher's Linear Discriminant Analysis (LDA) for high-dimensional data classification. By addressing the challenge of varying variances among features, we have successfully overcome a critical limitation of traditional LDA methods that assume uniform variances. Our approach, based on Nonparametric Maximum Likelihood Estimation (NPMLE) techniques, not only accommodates distinct variances but also demonstrates exceptional performance across a range of variance distribution profiles, including left-skewed, symmetric, and right-skewed forms.
Our empirical experiments have provided strong evidence of the practical benefits of our approach. We have shown that our methodology excels in accurately classifying high-dimensional data characterized by heterogeneous variances, contributing to the advancement of discriminant analysis techniques in high-dimensional settings. 
{Furthermore, we validated the effectiveness of our methodology through 
comprehensive analyses on various real datasets including gene expression datasets.}
These findings open up new opportunities for improving classification accuracy in a variety of real-world applications.
Of course, our research is based on the assumption of the Independent Rule (IR), but it is recognized that this assumption may not always be reasonable. Therefore, as part of future work, we aim to extend our proposed Multiple Variable Analysis (\textit{MVA}) by relaxing the IR assumption. Additionally, exploring the extension of \textit{MVA} to address classification problems in multiclass scenarios or dealing with class-imbalanced cases presents an intriguing avenue for further investigation.
}

\section*{Acknowledgement}
Seungyeon Oh and Hoyoung Park were supported by the National Research Foundation of Korea (NRF) grant funded by the Korea government(MSIT) (No. RS-2023-00212502). 
\bibliographystyle{apalike} %
\bibliography{refs} 

\end{document}